\documentstyle[12pt,aaspp4,epsfig]{article}

\journalid{337}{15 January 1989}
\articleid{11}{14}
\def\simlt{\mathrel{\hbox to 0pt{\lower 3.5pt\hbox{$\mathchar"218$}\hss}
      \raise 1.5pt\hbox{$\mathchar"13C$}}}
  \def\simgt{\mathrel{\hbox to 0pt{\lower 3.5pt\hbox{$\mathchar"218$}\hss}
      \raise 1.5pt\hbox{$\mathchar"13E$}}}

\begin{document}

\title{Inside-Out Infall Formation of Disk Galaxies: Do Predictions
Differ from Models without Size Evolution?}

\author{Rychard Bouwens} 
\affil{Department of Physics,,
    University of California,
    Berkeley, CA 94720}
\author{Laura Cay\'on}
\affil{Instituto de F\'\i sica de Cantabria,
       Universidad de Cantabria,
       Spain}
\centerline \&
\author {Joseph Silk}
\affil{Departments of Astronomy and Physics,   and Center for Particle
    Astrophysics, University of California,
    Berkeley, CA 94720}

\begin{abstract}

We develop an idealized inside-out formation model for disk galaxies
to include a realistic mix of galaxy types and luminosities that
provides a fair match to the traditional observables.  The predictions
of our infall models are compared against identical models with
no-size evolution by generating fully realistic simulations of the
HDF, from which we recover the angular size distributions.  We find
that our infall models produce nearly identical angular size
distributions to those of our no-size evolution models in the case of
a $\Omega = 0$ geometry but produce slightly smaller sizes in the case
of a $\Omega = 1$ geometry, a difference we associate with the fact
that there is a different amount of cosmic time in our two models for
evolving to relatively low redshifts ($z \approx 1-2$).  Our infall
models also predict a slightly smaller (11\% - 29\%) number of large
(disk scale lengths $> 4 h_{50} ^{-1}$ kpc) galaxies at $z \approx 0.7$ for the
CFRS as well as different increases in the central surface brightness
of the disks for early-type spirals, the infall model predicting an
increase by 1.2 magnitudes out to $z \approx 2 \ (\Omega = 0), 1\
(\Omega = 1)$, while our no-size evolution models predict an increase
of only 0.5 magnitude.  This result suggests that infall models could
be important for explaining the 1.2-1.6 magnitude increase in surface
brightness reported by Schade et al.\ (1995, 1996a, 1996b).

\end{abstract}


\section{Introduction}

Infall formation of disk galaxies predicts an inside-out formation
process that produces galaxy sizes smaller than expected in models
without intrinsic size evolution (Cay\'on, Silk \&
Charlot\markcite{cayon} 1996).  Prantzos \& Aubert\markcite{pa} (1995)
present a detailed comparison of several star formation models in an
effort to explain the observed properties of the Galactic disk. The
favoured model is one in which the star formation rate (SFR) is
proportional to the gas surface density and to the differential
rotation rate and in which gas evolution is driven by infall of
unenriched gas during a certain period of time in the life of the
galaxy (Wang \& Silk\markcite{wang} 1994).  In a first attempt to test
infall formation of field disk galaxies, Cay\'on et
al.\markcite{cayon} (1996) noted that
the observed trend toward small sizes at faint magnitudes and high
redshifts is clearly reproduced by these models.

In this Letter, we extend the development of the idealized infall
formation model for disk galaxies by incorporating a range
of different galaxy types and luminosities given by a slightly refined
version of the Pozzetti, Bruzual, \& Zamorani\markcite{poz} (1996)
pure luminosity evolution (PLE) model.
We aim to determine the extent to which these models
differ from models with no-size evolution in the prediction of the
angular size distributions, the number of large-intermediate redshift
disks, and the evolution in surface brightnesses observed in disk
galaxies at higher redshifts (Schade et al.\markcite{schade}
1995,\markcite{schadea} 1996a, \markcite{schadeb} 1996b; Simard \&
Prichet 1997\markcite{simpri}).
All calculations are performed with $H_o=50$ km/sec/Mpc 
and expressed in Vega magnitudes unless otherwise noted.

\section{Description of Model}

Infall formation models have been favored by studies of the
disk of our own Galaxy
(Ferrini et al.\markcite{ferrini} 1994;
Dopita \& Ryder\markcite{dopita} 1994; see Prantzos \& Aubert\markcite{pa} 
1995 for a detailed comparative analysis) and nearby galaxies (Ryder \&
Dopita\markcite{ryder} 1994). 
A model for disk galaxies based on infall formation
was presented in Cay\'on, Silk \& Charlot\markcite{cayon} (1996). Only a
brief description 
is included here.
A phenomenological model satisfying the observational requirements is
one with infall in which the SFR depends on radius and time following a
Schmidt-type law
\begin{equation}
SFR(r,t)=(1-R)^{-1}{{\Sigma_g(r,t)}\over {\tau_g(r)}}~~~
{\rm M}_{\odot} {\rm pc}^{-2} {\rm Gyr}^{-1},
\end{equation}
\noindent
where $R\approx 0.32$ is the returned fraction of mass that was formed
into stars, $\Sigma_g(r,t)$ is the gas surface density at radius $r$
and age $t$, and $\tau_g(r)=[0.3(1-R)(r_{\odot}/r)]^{-1}$~Gyr is the
gas consumption timescale ($r_{\odot }=8.5$ kpc).  Normalizing the
infall rate density of metal-free gas to the total surface density
$\Sigma_{tot}$ of stars plus gas observed at age $T$ Gyr
(corresponding to $z\sim 0$) and using the instantaneous recycling
approximation (IRA), the evolution of the gas surface density is given
by
\begin{equation}
\Sigma_g(r,t)=\Sigma_{tot}(r,T){{exp[-t/\tau_g(r)]-exp[-t/\tau_f(r)]}\over
 {\{1-[\tau_f(r)/\tau_g(r)]\}\{1-exp[-T/\tau_f(r)]\}}},
\end{equation}
\noindent where $\tau_f$ is the infall time scale.
As in our previous work, for
galaxies with Milky-way luminosities, i.e., $M_{b_j} \approx -20$, we
take $\Sigma_{tot}$ to equal the sum of the observed gas and the
stellar surface density necessary to yield the assumed ($z \approx 0$)
surface brightness profile.  For galaxies of different luminosities,
however, we take the star formation profile to be equal to the star
formation profile for Milky-Way galaxies, but stretched by an
appropriate factor in radius to produce the luminosity in question.

Closely following the Pozzetti et al.\ (1996) PLE model, we develop a
PLE model in terms of the above formalism for both an open and a flat
universe, where we use three different morphological types:
ellipticals, early-type spirals, and late-type spirals.  To improve
the agreement of our model with the number counts observed in the HDF
for an open geometry, we increase the normalization and steepen the
slope of our luminosity function slightly over the values given by
Pozzetti et al.\ (1996).  With these changes, we find that our models
fit the number counts, redshift distributions, and luminosity
functions as well as does the Pozzetti et al.\ (1996) model.

We tune both the age $T$ and the gas infall time scale $\tau_f$ to
match the $z=0$ colours and to provide a rough match with the star
formation histories given for the Sabc and Sdm morphological types
specified in the Pozzetti et al. (1996) PLE model with exponentially
decreasing star formation.  We take late-type spirals to have pure
exponential profiles, and early-type spirals to have an exponential
bulge added to this profile with a scale length that is 0.086 that of
the disk (Courteau, de Jong, \& Broeils 1996\markcite{courteau}) with
bulge-to-total ratios similar to those given in King \& Ellis
(1985)\markcite{ke85}.  At redshift zero, we assume that our disks
follow Freeman's law, i.e. $\mu_{bj} = 21.65$ $\textrm{mag/arcsec}^2$
(Freeman 1970\markcite{freeman}).  For galaxies with $M_{bj} > -18$,
however, we take the entire galactic population to be composed of
late-type exponential disks with $\mu_{bj} = 22.5$ to account for the
observed lack of early-type galaxies at low luminosities (Binggeli,
Sandage, \& Tammann 1988\markcite{bst}) and their lower surface
brightnesses (McGaugh \& de Blok 1997\markcite{mcg}).  For
ellipticals, we adopt the size-luminosity relationship given by
Binggeli, Sandage, \& Tarenghi (1984)\markcite{bst84}.  To perform the
calculations for the described model, we break up our disk galaxies
into approximately 30 annuli, on which we perform the spectral
synthesis for each using a recent version of the Bruzual \& Charlot
tables (Leitherer et al.\ 1997\markcite{lei}).  We summarize our
choice of parameters in Table 1.

We compare this model, hereinafter referred to as the ``Infall''
model, with a nearly identical ``No Infall'' model, in which there is
no evolution in the scale length on which star formation occurs, even
though the star formation history for each morphological type in this
model is the same as for the ``Infall'' model.  Furthermore, we ensure
that the surface brightnesses in the $bj$ band for both models are
exactly the same at $z = 0$.

\section{Evolution in Central Surface Brightness and Size}

For our first set of comparisons, we look at the extent to which the
``Infall'' models differ from ``No Infall'' models in terms of the
evolution of different morphological types in the $\mu _o$ -- $r_{hl}$
plane.  Because of the presence of the bulge in the center of the
early-type spirals, we take the central surface brightness $\mu_o$ to
equal the surface brightness extrapolated assuming an exponential disk
profile from the surface brightnesses at $\approx 0.5$ and $\approx
2.0$ disk scale lengths--a procedure that is intended to be an
approximate estimate of the central surface brightnesses that would be
determined through the multicomponent fits (Schade et al.\ 1996a).
Values in this plane are presented in Figure 1 for the models
considered, from redshifts corresponding to an age of 1 Gyr down to
$z\sim 0$.  The values for $L^*$ Sab-Sbc and Scd-Sdm galaxies for the
``Infall'' model are shown as solid lines in panels (a) and (b),
together with the predictions one magnitude above and below (shaded
areas). Brightening of approximately 1.2 magnitudes occurs for early
types, from $z=0$ to $z\sim 2 (\Omega =0.0), 1 (\Omega =1.0)$, the
rate of brightening being slower in the more open geometry due to the
longer time between the formation of the galaxy and $z \approx 1$.
These values would be in agreement with recent observations (Schade et
al.\markcite{schade} 1995,\markcite{schadea} 1996a,
\markcite{schadeb}1996b) if early-types are dominating the samples.
The faster the evolution in the central parts of the disk (larger
$T/\tau _f$ corresponding to earlier type disks) the larger is the
amount of brightening observed.  By contrast, our ``No Infall''
models, presented in panels (c) and (d) of Figure 1, show no such
differential evolution, so consequently the early-type galaxies in
these models produce only $\sim 0.5$ magnitude compared to the $\sim
1.2$ magnitude of brightening observed in the ``Infall'' models.
These results suggest that inside-out formation scenarios, such as
those we sketch in this paper, would not only make a difference in the
observations relative to scenarios with no-size evolution, but they
also could play an important role in the apparent surface brightness
evolution observed to $z \sim 1$.

As is clear from Figure 1, galaxies in our ``Infall'' models evolve
not only in surface brightness but in size relative to ``No Infall''
models.  In light of the present compilation of size information for
the CFRS (Lilly et al.\markcite{lillyf} 1997), we make a simple
estimate for the expected decrease in number of large galaxies
predicted in the CFRS at intermediate redshift on the basis of our
``Infall'' models relative to our ``No Infall'' models.  We find that,
for the CFRS selection criteria ($17.5 < I_{AB} < 22.5$), there are
807 (852) large (disk scale lengths $> 4$ kpc) galaxies per redshift
interval per $\textrm{degree}^2$ at $z \approx 0.7$ for our $\Omega =
0$ ($\Omega = 1)$ infall models compared to 719 (609) for our $\Omega
= 0$ ($\Omega = 1$) ``No Infall'' models, a decrease of 11\% ($\Omega
= 0$) to 29\% ($\Omega = 1$).

Finally, we examine the differences in the angular size distributions
produced by the models in the context of the HDF, the deepest and
highest resolution optical observation to date.  To this end, we
generate detailed images with areas equal to roughly 3.8 times the
area of the three WF chips and with properties which closely match
those observed in the F814W HDF images.  In our simulations, we place
galaxies 
with random orientation angles and no extintion
on the mock images assuming no correlation, add Poissonian
noise, smooth over a Gaussian kernel with a 0.06-arcsec sigma, and
then degrade the image by adding some noise.  After adding this noise,
we smooth the images again over a Gaussian kernel with a 0.02-arcsec
sigma to reproduce approximately the drizzled noise properties of the
images (Williams et al.\ 1996\markcite{wil}).  We use SExtractor 1.2b5
(Bertin \& Arnouts 1996\markcite{ber96}) to make a catalogue and to
determine both the magnitude (SExtractor's MAG\_BEST) and the
half-light radii\footnote{We take the half-light radius to equal the
radius of the aperture which contains half the light as determined by
SExtractor's best estimate of its total (MAG\_BEST).} of the objects in
the F814W images for each of our models as well as the F814W images
(version 2) of the three WF chips (1634 objects total).  For a
detection, we require an object, after smoothing with the PSF, to be
composed of at least 10 contiguous pixels at 2$\sigma$ above the
background.

As was our goal, in Figure 2 we demonstrate that the $\Omega = 0$
models provide fair fits to the HDF number counts, decomposed into
three different bins of half-light radii (though we are 20-40\% low in
the faintest magnitudes bins).  We then compare the angular size
distributions recovered from the simulations for our ``Infall'' models
with those recovered from both our ``No Infall'' models and the HDF in
Figure 3, where we decompose our angular size distribution into its
constituent morphological types.  We find that while both our $\Omega
= 0$ models are in fair agreement with the angular size distributions
found in the HDF, they also appear to be in fair agreement with each
other, even for the early-type galaxies which evolve so very
differently in the two models (Figure 1).  For our $\Omega = 1$
models, however, there is a clear difference between the angular size
distributions for the ``Infall'' and ``No Infall'' models,
particularly among the early-type spirals.

To understand this difference, it is important to note that the
profiles of galaxies in our ``Infall'' models tend to approach those
in our ``No Infall'' models over time, and so the convergence begins
in the center and grows outward.  Fundamentally, both our $\Omega = 0$
models produce similar distributions of recovered half-light radii
because at any redshift only those central parts of the galaxy for
which the profiles in the two models have started to converge can be
seen, these radii being small at high redshifts and large at low
redshifts as a result of $(1+z)^4$ cosmological dimming.  Galaxies in
our $\Omega = 1$ models, however, have less time to evolve to
relatively low redshifts where cosmological dimming is less
significant (note that in Figure 1 the redshift corresponding to the
peak in surface brightness is lower for our $\Omega = 1$ models);
therefore, the apparent differences between the profiles produced by
the two models can be more readily observed.  In Figure 3, we plot the
distribution of true half-light radii (calculated from our code which
placed each galaxy on our simulated images) for the two $\Omega = 0$
models to illustrate the reality of the actual but relatively
unobservable differences between these two models.

In summary, we have extended the idealized infall model for disk
formation to include a realistic mix of galaxy types and luminosities
which provide a rough match to the number counts, redshift
distributions, and angular size distributions in the HDF.  We compare
the predictions of these models against identical models with no-size
evolution to see how our infall models would affect the observations.
For these infall models, we find that the number of large disks (disk
scale lengths $> 4$ kpc) is 11\%-29\% smaller than for our no-size
evolution models at $z \approx 0.7$ for the CFRS.  We also find
increases in the central surface brightness of disk galaxies of about
$1.2$ $\textrm{mag/arcsec}^2$ out to $z \approx 1 (\Omega = 1), 2
(\Omega = 0)$ for early-type spirals, very different from the $0.5$
$\textrm{mag/arcsec}^2$ increase found for models with no-size
evolution.  Our finding strongly suggests that inside-out formation
scenarios for disk galaxies should have observable consequences at high 
$z$ and
could be important for understanding the observed increase in surface
brightness of disks out to $z \approx 1$.

For our $\Omega = 0$ models, we find no significant differences
between the recoverable angular size distributions predicted on the
basis of our infall models and those predicted on the basis of our
models with no-size evolution.  We suspect that significant
differences between these models would have been obtained if the
early-type galaxies had been allowed to begin forming at lower
redshifts, as we found in the case of our $\Omega = 1$ models.
However, in this case, there would have not been as many galaxies at
high redshifts and our models would have underpredicted the number of
galaxies at faint magnitudes.  We therefore suspect that for any
simple single-formation-epoch PLE models which fit
the number counts, infall and no-size evolution models will tend to
produce very similar angular size distributions given the current
state of observations.  Therefore, if our infall models are correct,
the fact that other workers (Ferguson \& Babul 1997\markcite{fb97})
have found angular size distributions with no-size evolution PLE
models which tend to agree with the HDF
observations is not very surprising in light of the present result.

\acknowledgments

We would like acknowledge Ian Smail for his suggested procedure for
the determination of half-light radii.  We would like to thank
Emmanuel Bertin for providing us with SExtractor and Stephane Charlot
for making the Bruzual/Charlot spectral synthesis curves available.
We acknowledge useful conversations with Steve Zepf and Harry
Ferguson.  L.C. thanks both the Center for Particle Astrophysics and
Berkeley Astronomy Department for their hospitality during her stay in
Berkeley.  R.J.B. gratefully acknowledges support from an NSF graduate
fellowship. This research has also been supported in part by grants
from NASA and NSF.

\newpage	

\begin{center}
\epsfig{file=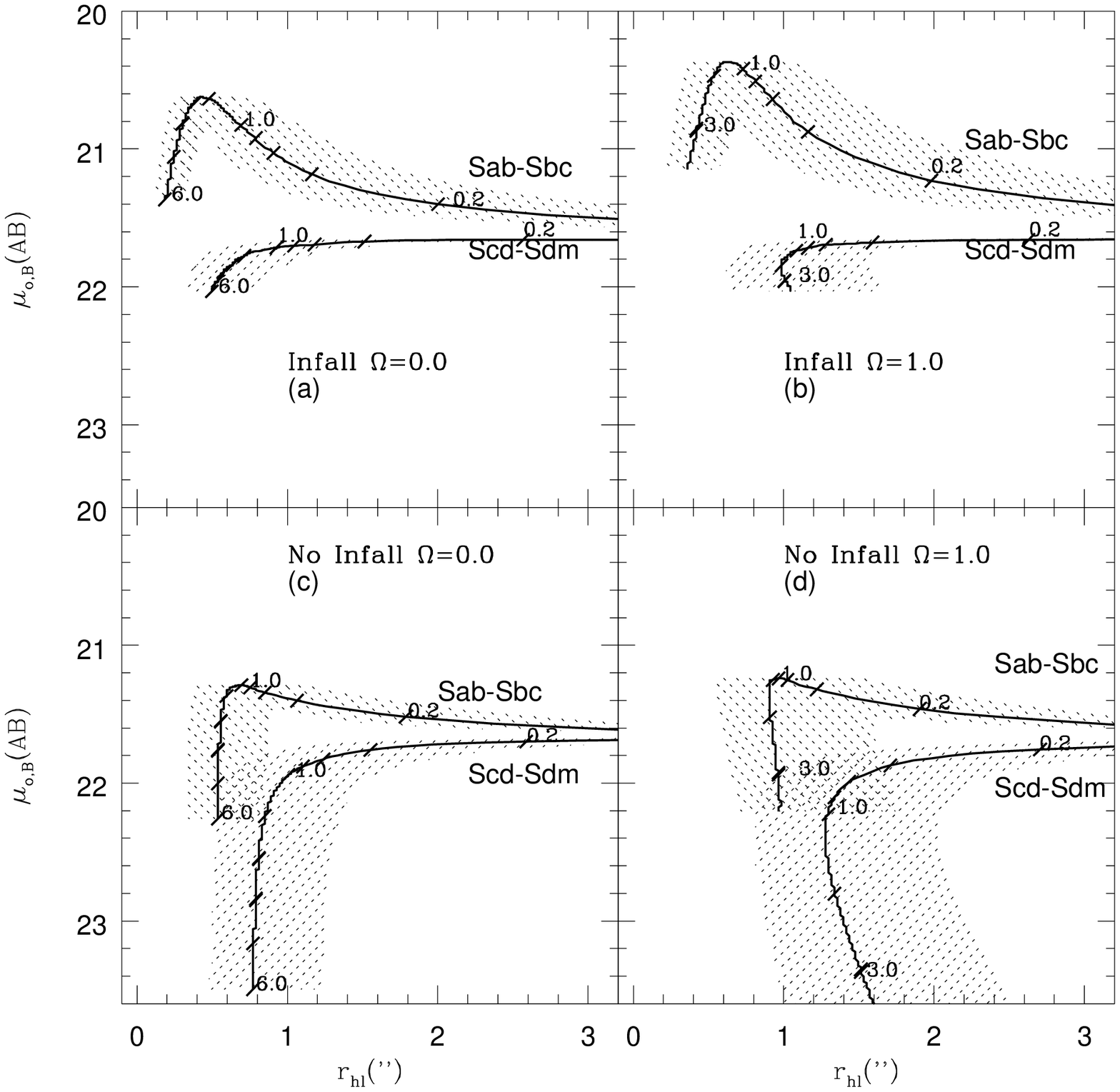,width=100mm}
\end{center}

\figcaption{Disk central surface brightness $\mu _o$ versus half-light
radius $r_{hl}$ for Sab-Sbc (top lines) and Scd-Sdm (bottom lines)
galaxies.  The redshift interval considered in the calculations
extends from $z=0$ up to a redshift that corresponds to the time at
which galaxies have been evolving 1 Gyr in the different models.
Solid lines denote the values corresponding to a $M^*$ galaxy with the
shaded areas covering one magnitude above and below this.  Predictions
of the ``Infall'' models are presented in panels (a) and (b) while the
``No Infall'' models are presented in panels (c) and (d). Predicted
values in $\Omega =0.0$ and $\Omega =1.0$ cosmologies appear in the
left panels and right panels respectively.  Tick marks denote the
locations at different redshifts in steps of $0.2$ from $z=0$ to $z=1$
and in steps of 1 from $z=1$ to the highest redshift indicated for
each model.}

\newpage

\begin{center}
\epsfig{file=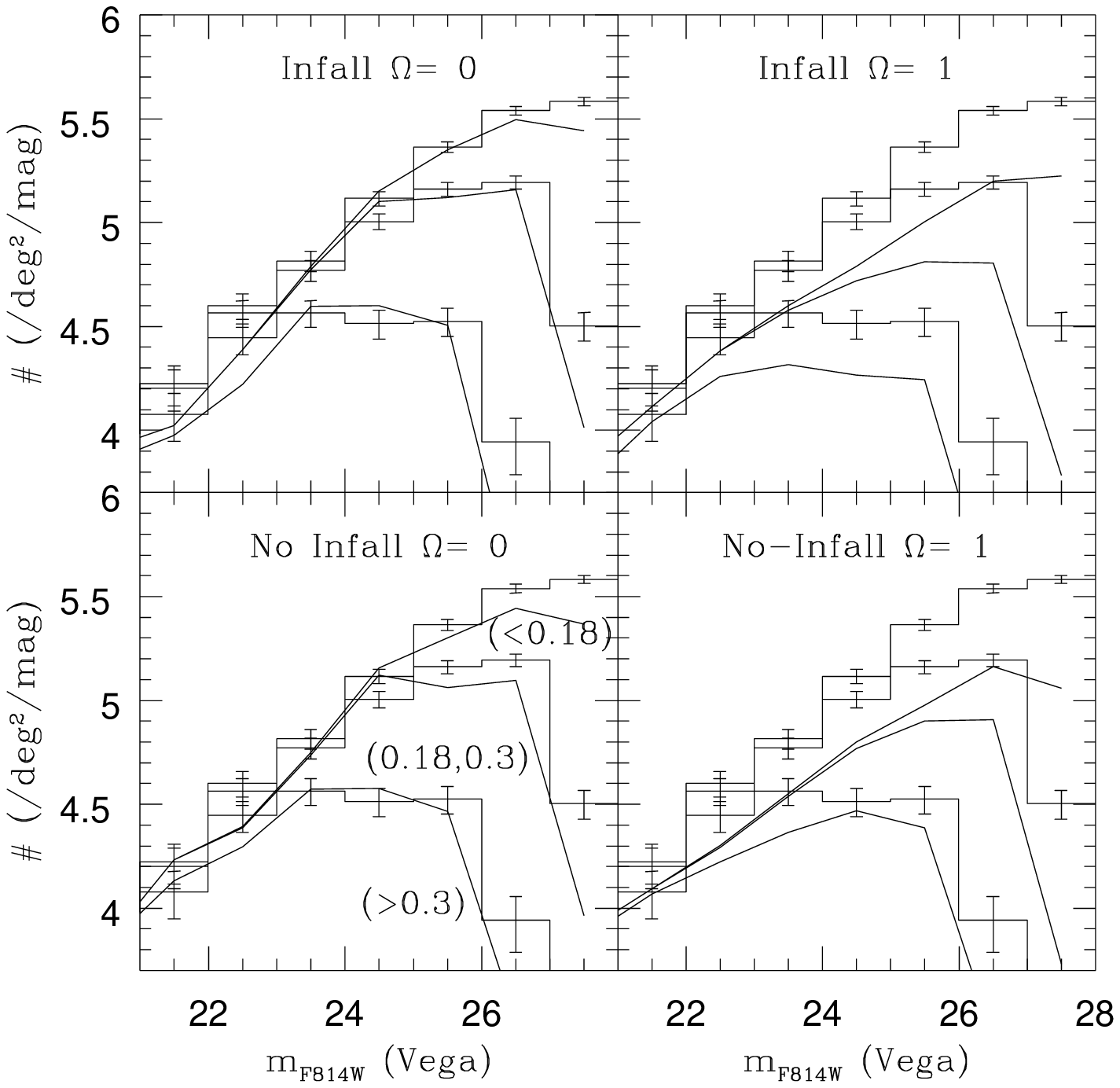,width=150mm}
\end{center}

\figcaption{ Comparison of the number counts recovered from the HDF
with those recovered from the simulations.  The observed number counts
are broken up in terms of their measured half-light radii into three
bins $r_{hl}<0.18$ $\textrm{arcsec}$, $0.18$ $\textrm{arcsec} < r_{hl} <
0.30$ $\textrm{arcsec}$, and $0.30$ $\textrm{arcsec} < r_{hl}$.  The error
bars are the one-sigma poissonian values.  }

\newpage

\begin{center}
\epsfig{file=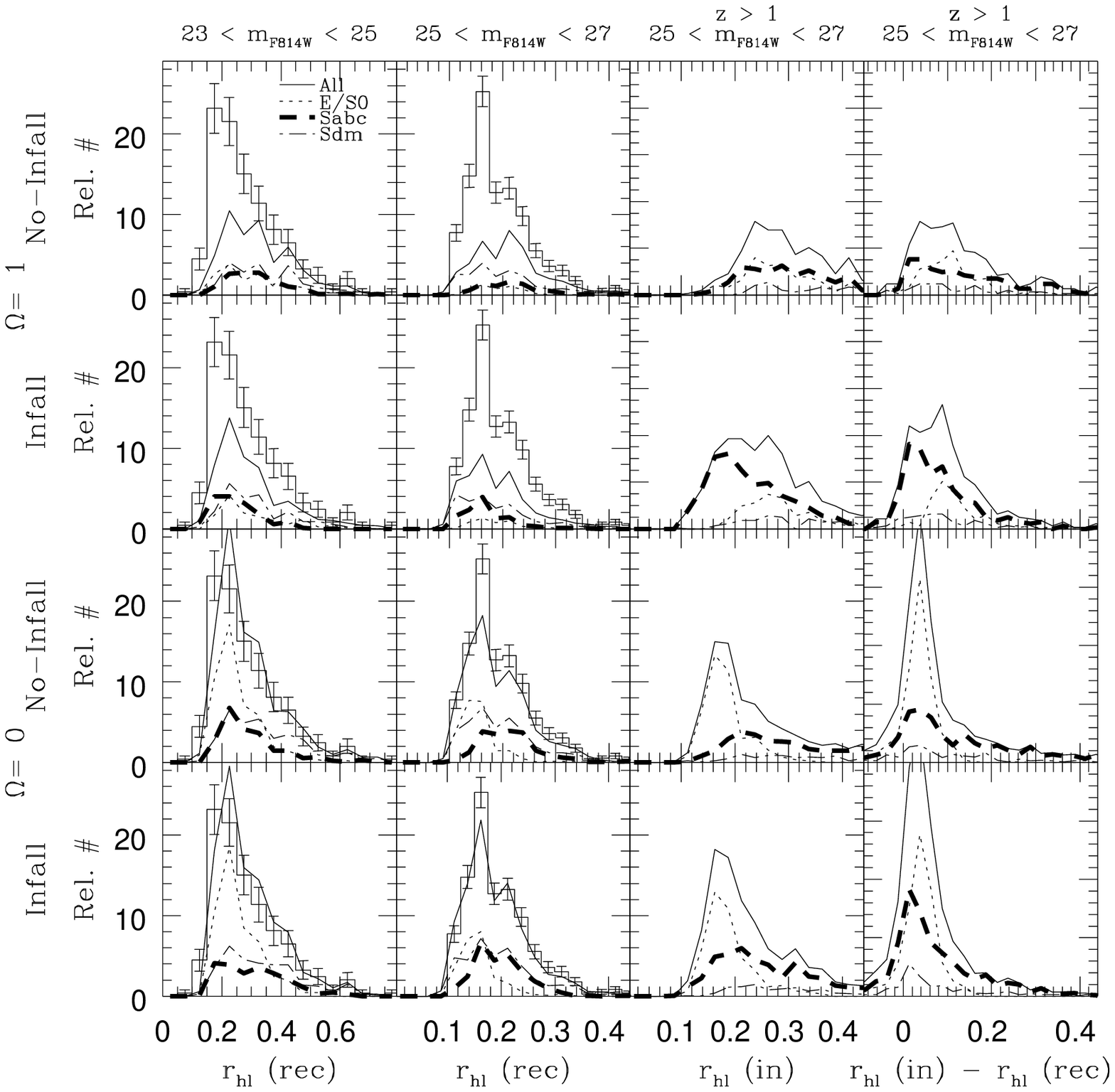,width=150mm}
\end{center}

\figcaption{ Comparison of the half-light radii distributions
recovered from the HDF (histogram) with one-sigma error bars against
those recovered ($\textrm{r}_{hl} \textrm{(rec)}$) from our
simulations in the magnitude bins ($23 < m_{F814W} \textrm{(Vega)} <
25$; $25 < m_{F814W} \textrm{(Vega)} < 27$).  The angular size
distributions recovered from the simulations are decomposed into
morphological types.  The true half-light radii ($\textrm{r}_{hl}
\textrm{(in)}$) distributions of higher redshift ($z > 1$) objects
recovered in the magnitude bin $25 < m_{F814W} \textrm{(Vega)} < 27$ are
compared to the recovered distributions to illustrate both the errors
in our procedure for determining the half-light radii and the fact
that differences in half-light radii between the ``Infall'' and ``No
Infall'' models are more apparent in the true half-light radii
distributions than in the recovered distributions.  }

\clearpage

\begin{deluxetable}{lrrrrrrrr}
\tablenum{1}
\tablewidth{0pt}
\tablecaption{Model parameters$^{a}$}
\tablehead{
\colhead{$\Omega $} &
\colhead{Type} & $\alpha$ & $M_{b_J} ^*$ & $\phi_0 ^*$ &
\colhead{$\textrm{B/T}^{b}$} & \colhead{$\mu_0 ^{b_J}$} &
\colhead{SFR}  &
\colhead{T(Gyr)}}
\startdata
0.0(1.0) & E/S0 & -0.48 & -20.87 & 0.95 & 1 & -- &$\tau _1^{c}$($B_1^{d}$)& 16.0
(12.7)\nl
\nodata & $\textrm{Sab-Sbc}^{e}$ & -1.32 & -21.14 & 0.69 & 0.2 & 21.65 & 
 $\tau _f=7.2$ Gyr(3.9 Gyr) & 18.0(12.7)  \nl
\nodata & $\textrm{Sab-Sbc}^{e}$ & -1.32 & -21.14 & 0.69 & 0.13 & 21.65 &
 $\tau _f=7.2$ Gyr(3.9 Gyr) & 18.0(12.7)  \nl
\nodata & $\textrm{Scd-Sdm}^{e}$ & -1.32 & -21.14 & 0.76 & 0 & 21.65 &
$\tau _f=\infty $ Gyr($\infty $ Gyr)& 18.0(12.7)  \nl
\nodata & $\textrm{Scd-Sdm}^{f}$ & -1.5 & -21.14 & 1.35 & 0 & 22.5 & $\tau
_f=\infty $ Gyr ($\infty $ Gyr)& 18.0(12.7)  \nl

\enddata
\tablenotetext{a}{Scalo IMF assumed for the E/S0 and Sab-Sbc types
and Salpeter IMF for Scd-Sdm types.} 
\tablenotetext{b}{Bulge-to-total luminosity ratio in $bj$ band at $z = 0$.}
\tablenotetext{c}{Exponential SFR characterized by decay time
$\tau _1=1$ Gyr.}
\tablenotetext{d}{1 Gyr burst SFR.}
\tablenotetext{e}{LF truncated to include only galaxies with $M_{bj} < -18$.}
\tablenotetext{f}{LF truncated to include only galaxies with $M_{bj} > -18$.}

\end{deluxetable}


\begin{references}


\reference{ber96} Bertin, E., \& Arnouts, S. 1996, \aaps, 117, 393.

\reference{bst84} Binggeli, B., Sandage, A., \& Tarenghi, M. 1984, \apj, 89, 64.


\reference{bst} Binggeli, B., Sandage, A., \& Tammann, G.A. 1988, \araa, 26, 509.

\reference{cayon} Cay\'on, L., Silk, J. \& Charlot, S. 1996, \apjl , 467, L53.

\reference{courteau} Courteau, S., de Jong, R.S. \& Broeils, A.H. 1996, \apjl , 
457, 
L73.

\reference{dopita} Dopita, M. \& Ryder, S. 1994, \apj , 430, 163.

\reference{ferrini} Ferrini, F., Molla, M., Pardi, M., \& Diaz, A. 1994, \apj , 
427, 745.

\reference{ke85} King, C.R., Ellis, R.S. 1985, \apj, 288, 456.

\reference{fb97} Ferguson, H.C., \& Babul, A. 1997, in preparation.

\reference{freeman} Freeman, K.C. 1970, \apj, 160, 811.

\reference{lei} Leitherer, C., et al.\ 1996, 108, 996.

\reference{lilly} Lilly, S.J., et al.\ 1997, in preparation.

\reference{mcg} McGaugh, S.S., \& de Blok, W.J.G. 1997, \apj, 481, 689.



\reference{poz} Pozzetti, L., Bruzual, G. \& Zamorani, G. 1996, \mnras , 281, 953.

\reference{pa} Prantzos, N. \& Aubert, O. 1995, \aap , 302, 69.


\reference{ryder} Ryder, S. \& Dopita, M. 1994, \apj, 430, 142.

\reference{simpri} Simard, L., \& Pritchet, C.J. 1997, astro-ph/9606006.

\reference{schade} Schade, D., Lilly, S.J., Crampton, D., Hammer, F., Le Fevre, O. 
\& Tresse, L. 1995, \apjl , 451, L1.

\reference{schadea} Schade, D., Lilly, S.J., Le Fevre, O., Hammer, F. \& Crampton, 
D. 1996a, \apj , 464, 79.

\reference{schadeb} Schade, D., Carlberg, R.G., Yee, H.K.C., L\'opez-Cruz, O. \& 
Ellingson, E. 1996b, \apjl , 465, L103.


\reference{was} Wang, B. Q. \& Silk, J. 1994, \apj, 427, 759.

\reference{wil} Williams, R.E., et al.\  1996, \aj, 112, 1335.

\end{references}
\end{document}